\documentstyle[graphicx,times]{jaa}

\begin{document}
\title[4.8 GHz Intra-Day Variability of FSRQ 0507+179]{4.8 GHz Intra-Day Variability of FSRQ 0507+179}
\author[L. Cui et al.]
        {Lang Cui$^{1,2,}$\thanks{e-mail: cuilang@xao.ac.cn},
        Jun Liu$^{1,2}$, \& Xiang Liu$^{1,2}$ \\
        $^1$Xinjiang Astronomical Observatory, Chinese Academy of Sciences,\\
        150 Science 1-Street, Urumqi, XinJiang 830011, China \\
        $^2$Key Laboratory of Radio Astronomy, Chinese Academy of Sciences,\\
        2 West Beijing Road, Nanjing, JiangSu 210008, China\\
        }
\maketitle
\label{firstpage}
\begin{abstract}
As one of targets of many flux monitoring campaigns, the FSRQ 0507+179 shows various flux variation
properties at almost all observing wavelengths, from radio to $\gamma$-ray. With Urumqi 25-m telescope,
our study on this object is focusing on its radio flux variability, especially the Intra-Day Variability (IDV).
We carried out a total of six epochs of IDV observations on 0507+179 at 4.8 GHz since Mar. 2010, found
clearly IDV behaviors in all observing sessions and consider it is likely a type I IDV source by analyzing
the characteristics and the timescales of the light curves. Additionally, we found 0507+179 exhibited
some different IDV behaviors after an optical flare.

\end{abstract}


\section{Introduction}
FSRQ 0507+179 is a typical blazar with the redshift of 0.416 (Perlman et al. 1998) and
was detected gamma-ray emission by Fermi/LAT(Abdo et al. 2010).
The 5 GHz VLBI observations reveal a compact core-jet structure with the linear
size of $\sim$100 pc (Cui et al. 2011). As one of targets of many flux monitoring programs,
the FSRQ 0507+179 shows various flux variation properties at almost
all observing wavelengths, from radio to $\gamma$-ray. To study rapid flux
variability in 0507+179 at radio band, we carried out a total of six IDV sessions at 4.8 GHz
with Urumqi 25-m telescope from 2010 to 2012. In following sections, we will
present the IDV observations and results and give a brief summary finally.

\section{Observations and Results}
Ever since Sep. 2008, we carried out a monthly flux monitoring program on 169 compact radio
sources with Urumqi 25-m telescope at 4.8 GHz. As one of the targets of the monitoring plan, 0507+179 was
found showing significant monthly flux variations during the term of Sep. 2008 to Nov. 2012, see Figure 1.
To find more rapid flux variability in 0507+179, we consequently carried out IDV observations at
the same frequency since Mar. 2010.

\begin{figure}
     \centering
     \includegraphics[width=\textwidth,height=7.8cm]{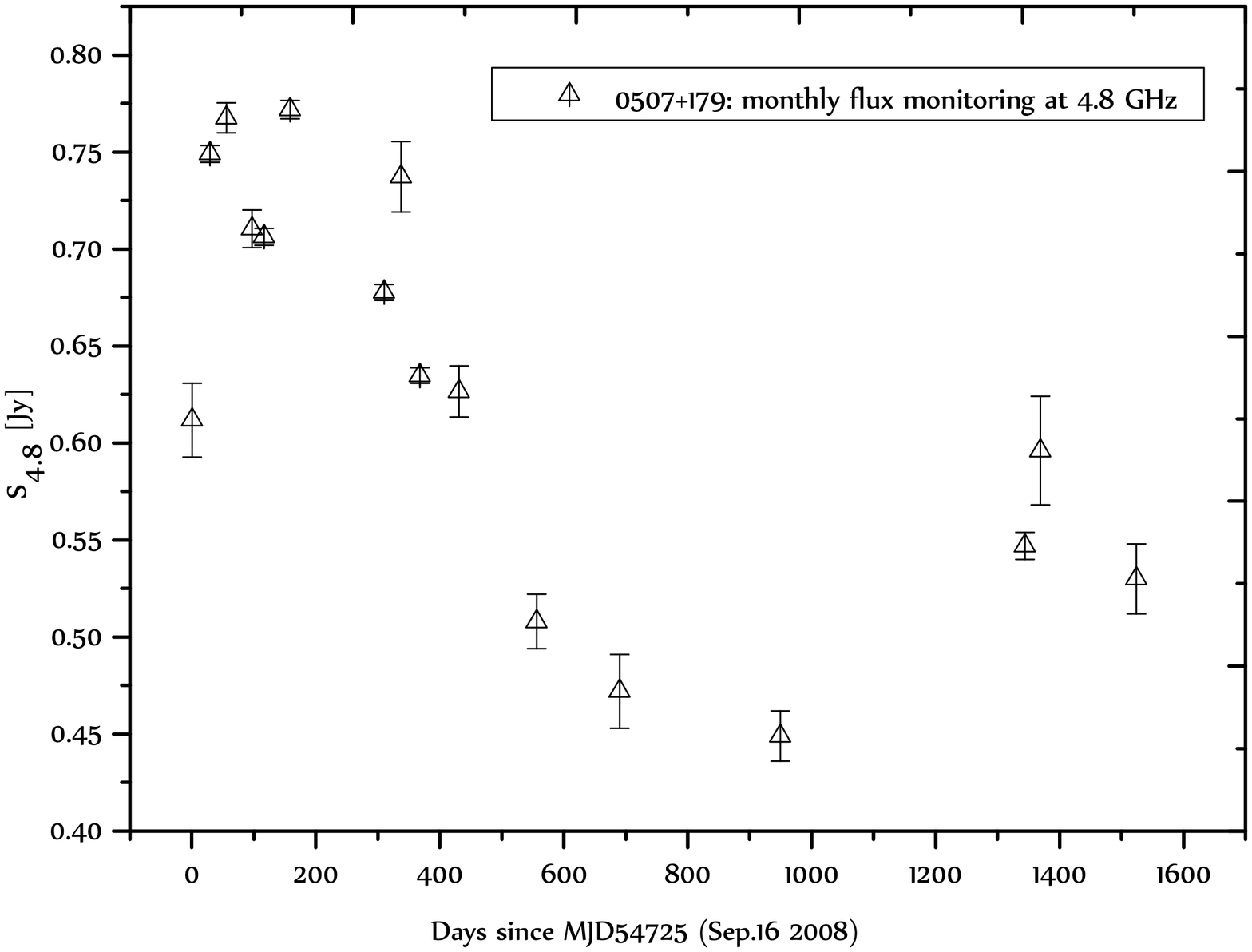}
     \caption[]{The light curve of monthly flux monitoring on 0507+179 at 4.8 GHz. }
   \end{figure}

The IDV observations were made in `cross-scan' mode and each scan consists of 8 sub-scans along azimuth
and elevation across the target position. The detailed data calibration procedure can be found in
Kraus et al. (2003) and Liu et al. (2012). The results of IDV observations on 0507+179 are given
in Table 1, which lists the epoch, the number of scans, mean value of flux density, the standard
deviation of flux density, the modulation index of calibrators, the modulation index of 0507+179, the
relative variability amplitude, the reduced Chi-Square and the timescale obtained from structure
function (SF) analysis (except the second and fifth ones, which were roughly estimated by eye according to
the light curves, since there are too few data points to obtain the timescales by using SF analysis).

In Sep. 2012, 0507+179 was detected an optical flare by the MASTER (Mobile Astronomical System of the TElescope-Robots),
and the magnitude increased fast from 18.7 to 16.6, upto the brightest magnitude since Jan. 2009 (ATel \#4424).
The last IDV session in Table 1 was arranged after the optical flare to study the possible difference of IDV characteristics
compared to that before. Then all the light curves of the IDV observations on 0507+179 at 4.8 GHz are
exhibited epoch by epoch in Figure 2.

\begin{table}
\caption{The results of IDV observations on 0507+179 at 4.8 GHz.}
\begin{center}
\begin{tabular}{ccccccccc}
\noalign{\smallskip}
\hline\hline
\noalign{\smallskip}
Epoch       &  N    & $<S>$  & $\Delta S$  & $m_{0}$   &   $m$   &  $Y$    &  $\chi_{r}^{2}$  & $t_{SF}$  \\
            &       & $[Jy]$ & $ [Jy]$     & $[\%] $   &  $[\%]$ & $[\%]$  &                  & $[d] $    \\
\noalign{\smallskip}
\hline
\noalign{\smallskip}
25.03.2010 &  15  &  0.508   &  0.014   &  0.80   &   2.84   &  8.16   &  6.982     &  2.1$\pm$0.2   \\
05.08.2010 &   9  &  0.472   &  0.019   &  0.90   &   4.11   &  12.04  &  8.859     &  $>$2.0     \\
21.04.2011 &  14  &   0.449  &  0.013   &  0.60   &   2.89   &  8.49   &  5.506     &  2.0$\pm$0.1  \\
18.05.2012 &  13  &   0.547  &  0.033   &  0.60   &   6.03   &  18.01  &  49.273    &  2.2$\pm$0.3    \\
15.06.2012 &   8  &   0.596  &  0.028   &  0.60   &   4.77   &  14.19  &  18.657    &  $>$2.0      \\
15.11.2012 &  25  &   0.530  &  0.018   &  0.60   &   3.34   &  9.85   &  17.469    &  1.2$\pm$0.2  \\
\noalign{\smallskip}
\hline
\end{tabular}
\end{center}
\end{table}

\begin{figure}
     \centering
     \includegraphics[angle=-90, width=0.475\textwidth]{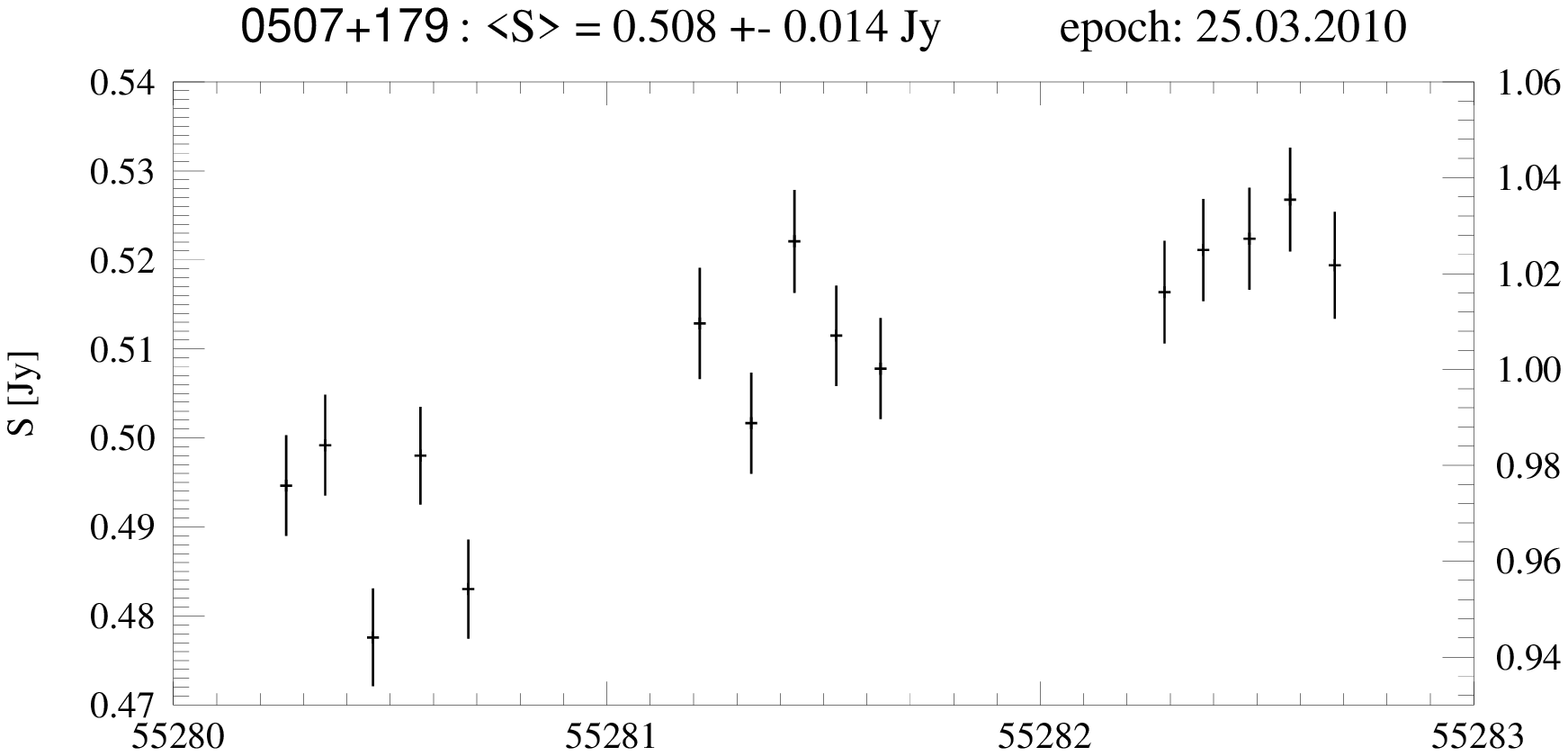}
     \includegraphics[angle=-90, width=0.475\textwidth]{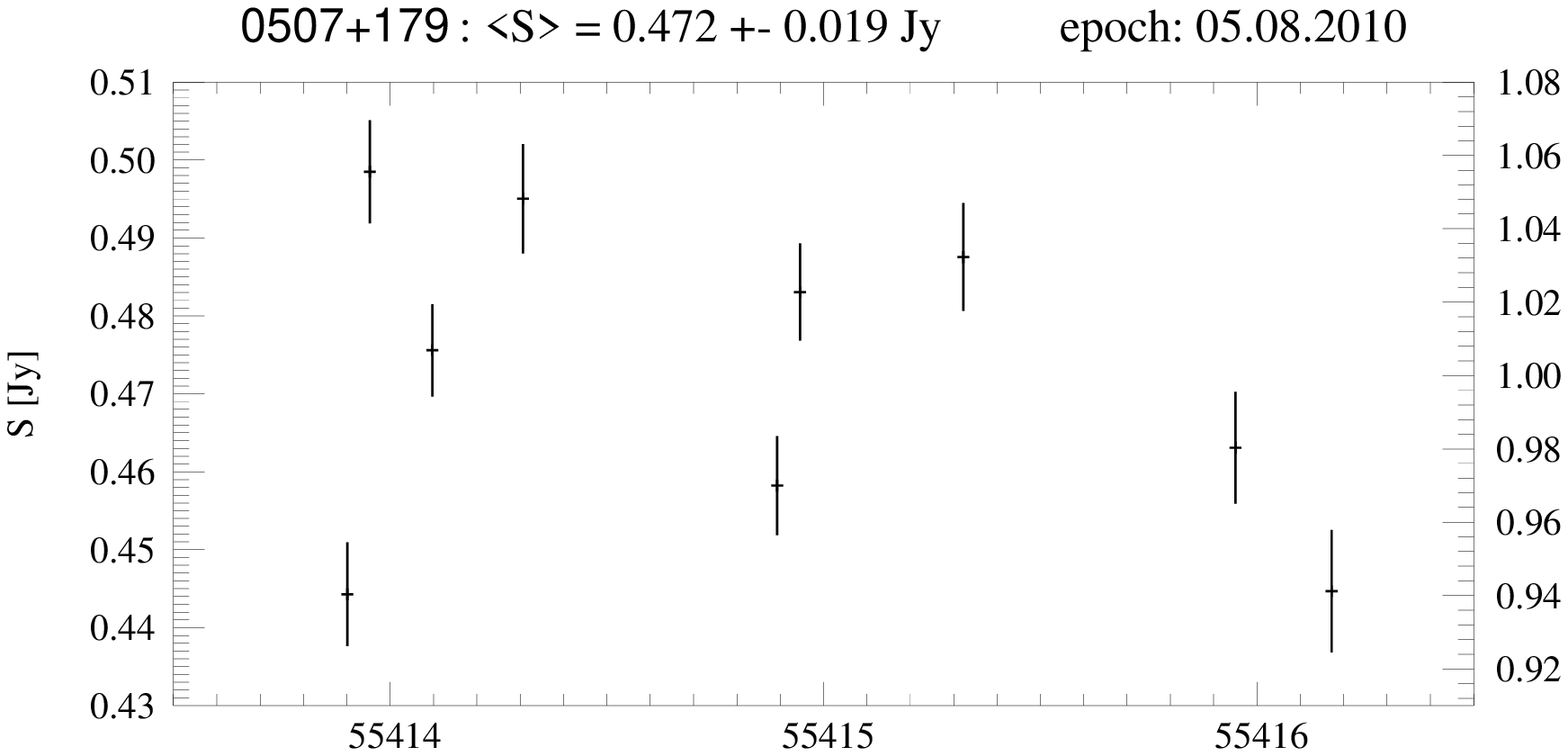}
     \includegraphics[angle=-90, width=0.475\textwidth]{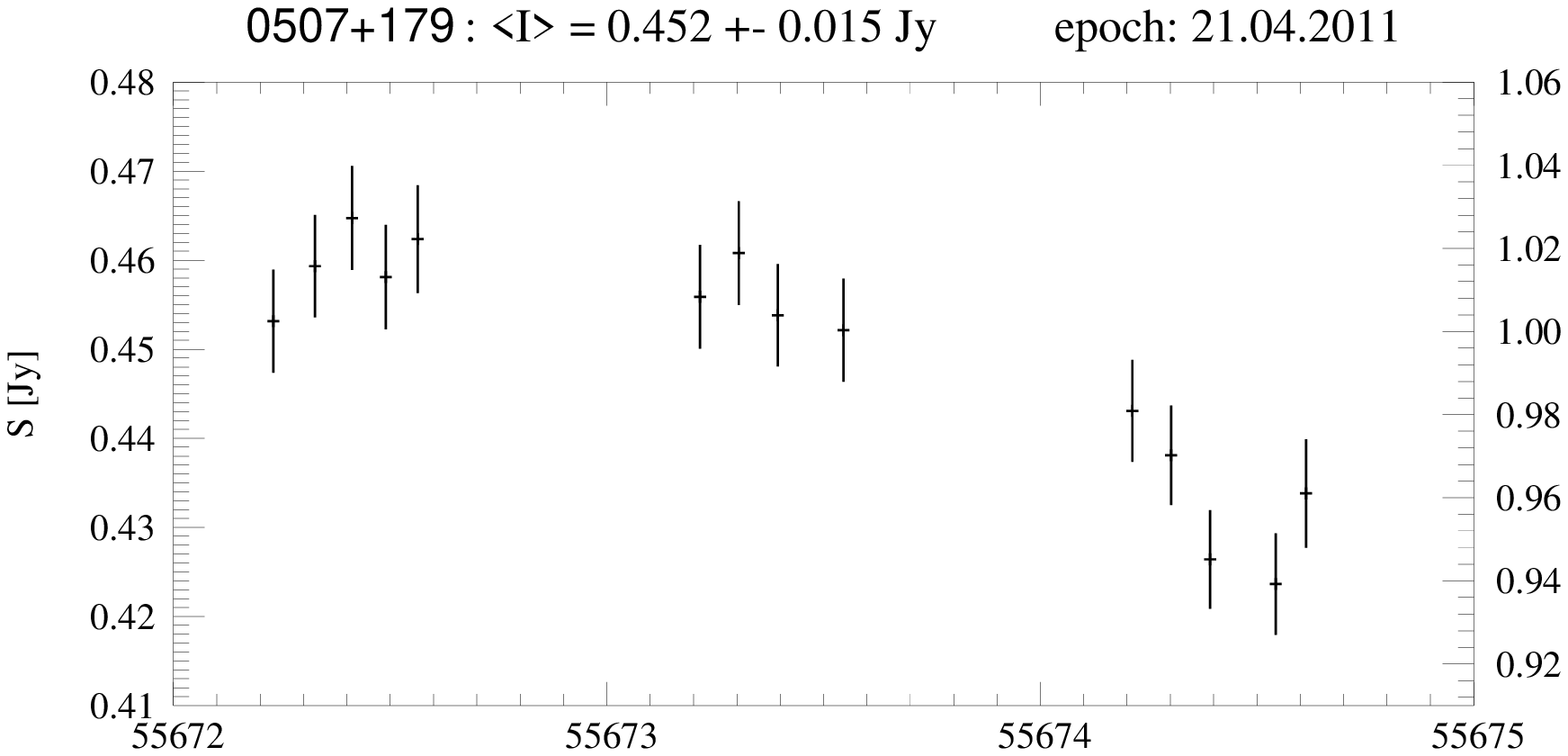}
     \includegraphics[angle=-90, width=0.475\textwidth]{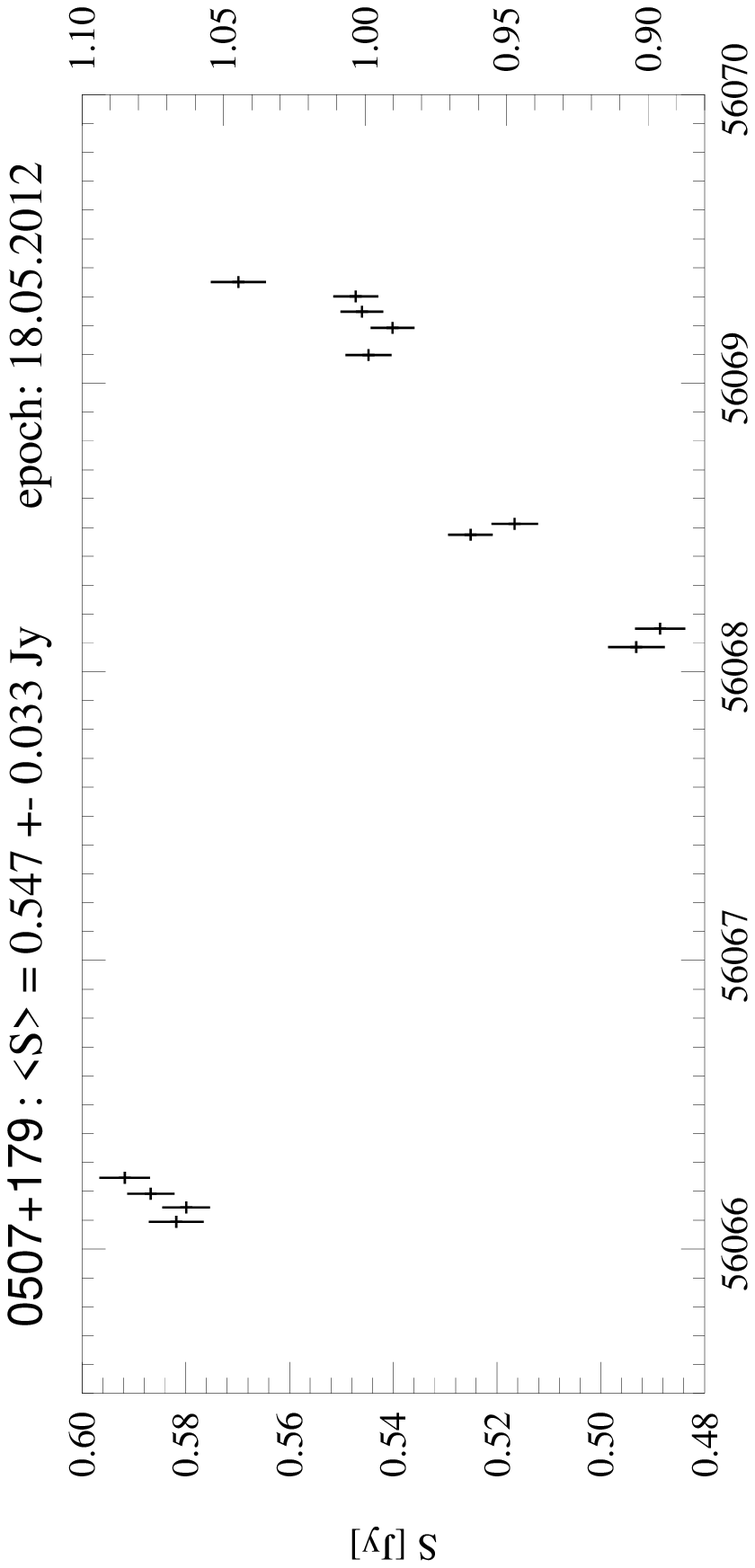}
     \includegraphics[angle=-90, width=0.475\textwidth]{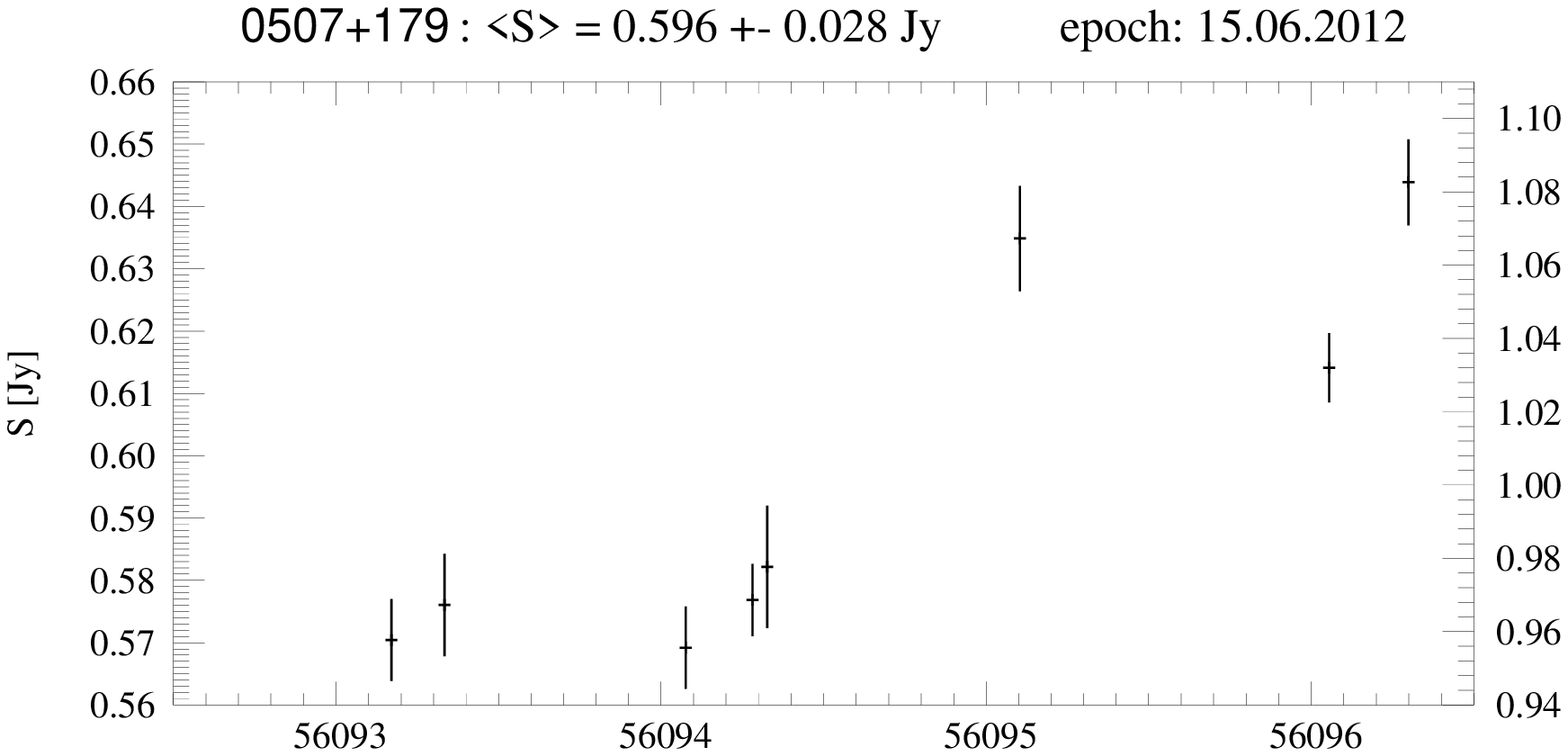}
     \includegraphics[angle=-90, width=0.475\textwidth]{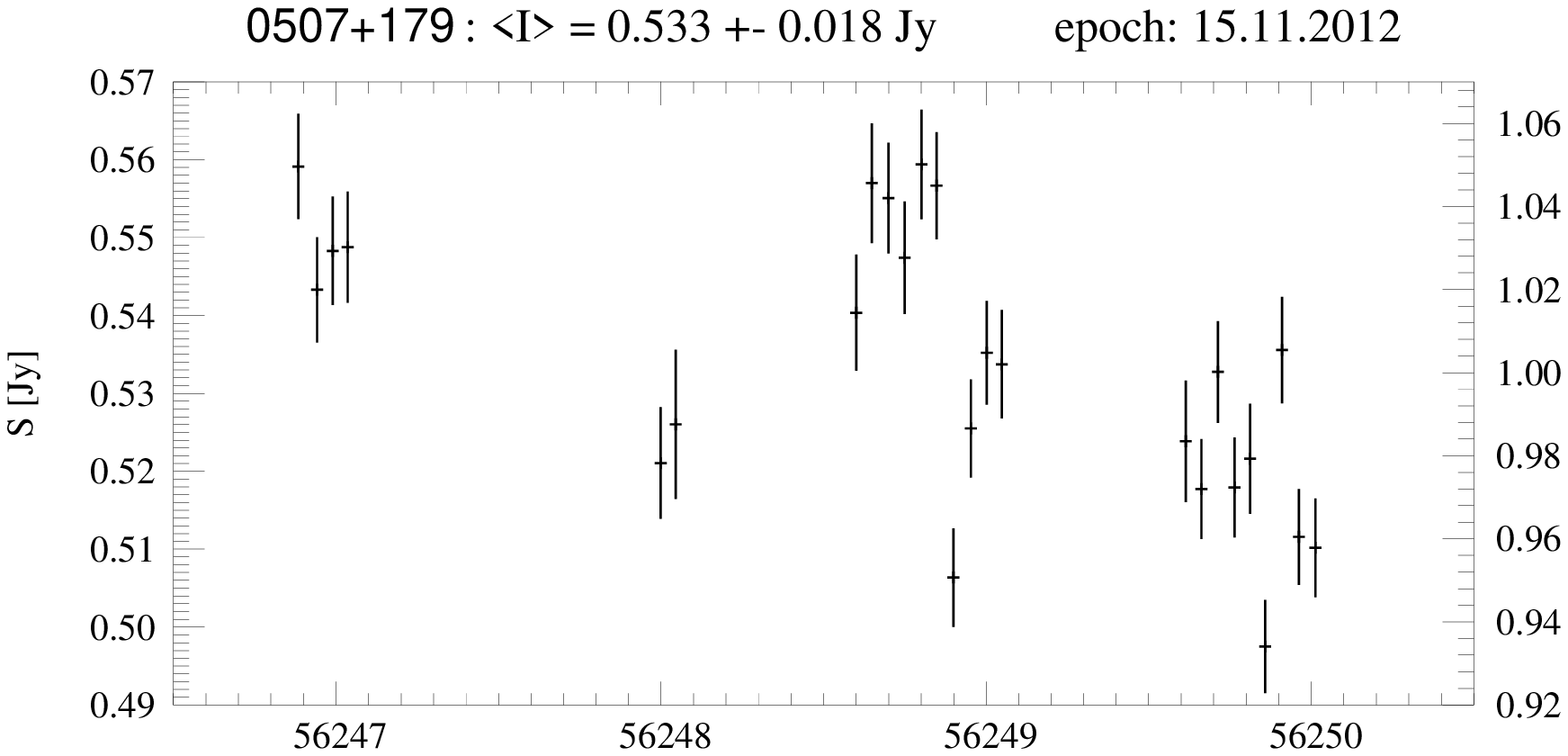}
     \caption[]{The light curves of IDV observations on 0507+179 at 4.8 GHz.}
   \end{figure}
\section{Discussion and Conclusion}
The meanings of the parameters which used to evaluate the significance and amplitude of
the variability listed in Table 1, namely the modulation index $m$, the relative variability amplitude $Y$ and the
reduced Chi-Square $\chi_{r}^{2}$ can be found in Quirrenbach et al. (2000) and Kraus et al. (2003). According
to a Chi-Square-test (Bevington \& Robinson 1992) with the reduced Chi-Square listed in Table 1,
0507+179 shows obvious IDV behaviors in all observing sessions at a confidence level of over 99.9\%.
And almost all the light curves except the last session in Figure 2 show a monotonically increasing or
decreasing trend, with the timescales of more than two days, which indicates that 0507+179 is
likely a type I IDV source (Quirrenbach et al. 2000).

To study the possible difference of IDV characteristics in 0507+179 after the optical flare detected by the MASTER,
we consequently arranged a new IDV session in the middle of Nov. 2012.
Compared to the other epochs before the optical flare in Figure 2, the light
curve of the last epoch becomes complex instead of monotonically increasing or decreasing and the timescale
becomes shorter than before, just as listing in Table 1. This implies that the radio flux variability of 0507+179
becomes more rapid and it may being turn into the high active state after the optical flare. Further radio
monitoring of the object is necessary to study its possible radio outburst.

In summary, we carried out six IDV sessions on 0507+179 with the Urumqi 25-m radio telescope at 4.8 GHz
since Mar. 2010 and found that the source exhibits clearly IDV behaviors in all observing sessions.
The relatively longer IDV timescales in most of the sessions indicate the source is likely a type I IDV source.
Additionally, we found 0507+179 exhibited some different IDV behaviors that the flux variability became more
rapid after an optical flare.
\\
\\
\textbf{Acknowledgements}
\\
This work is supported by the program of
the Light in China's Western Region (LCWR) under grant XBBS201024,
the National Natural Science Foundation of China under grant
11073036 and the 973 Program of China (2009CB824800).

\label{lastpage}
\end{document}